\documentclass[11pt,a4paper]{article}

\usepackage[margin=2cm]{geometry}
\usepackage{amsmath,amssymb,amsthm}
\usepackage{graphicx}
\usepackage{authblk}
\usepackage[hyphens]{url}
\usepackage[numbers]{natbib}
\usepackage[hidelinks]{hyperref}
\newcommand{\captionwindows}{Positions of the fixed and tracked windows in the three videos. (a) The Hualien video, with a single fixed window of 40$\times$80 px because rotation, panning, and blur are minor. (b), (c) The Mandalay and Kumamoto videos, with two fixed windows of 50$\times$50 px. In each panel, all analyzed frames are superposed with transparency after the shift, scaling, and rotation that keep the fixed windows aligned; the ground near the cameras therefore appears stationary.}
\newcommand{\captionHualien}{Results for the video of the 2018 Hualien earthquake. (a) Displacement trajectory; color indicates the cross-correlation coefficient between the reference window and each frame. (b) Displacement histories in the $x$ and $y$ directions, with the same color scale as in panel (a). (c) Estimated velocity history. The time intervals are unequal because the original and captured videos have different frame rates.}
\newcommand{\captionMandalay}{Same as Fig.~\ref{fig:Hualien}, but for the 2025 Mandalay earthquake.}
\newcommand{\captionKumamoto}{Same as Fig.~\ref{fig:Hualien}, but for the 2026 Kumamoto earthquake. Although some outliers are shown in gray in panels (a) and (b), they have little effect on the velocity estimate in panel (c) because observations with low correlation coefficients are assigned lower weights.}

\title{Co-seismic surface fault displacement captured in videos: image tracking across three earthquakes}

\author[1]{Shiro Hirano\thanks{Corresponding author: \texttt{hirano@hirosaki-u.ac.jp}}}
\affil[1]{Hirosaki University, 3 Bunkyo, Hirosaki, Aomori 036-8561, Japan}

\begin{document}

\maketitle

\begin{abstract}
    Surface displacements, including co-seismic fault slip, have occasionally been captured by closed-circuit television (CCTV) cameras. Such video recordings may provide information unavailable from seismometers located away from the fault. We analyzed videos of surface displacement from three earthquakes---the 2018 Hualien earthquake, the 2025 Mandalay earthquake, and the 2026 Kumamoto earthquake---using a unified methodology. We removed the effects of strong ground motion by aligning each frame so that reference regions in the image remained fixed. This analysis enables the temporal evolution and, in particular, the duration of surface displacement to be estimated. Although the three earthquakes ranged from $M_\mathrm{w}$6.4 to $M_\mathrm{w}$7.7, the tracked motion lasted approximately 2 s in all three cases; in the Hualien case, the motion included compression of the ground rather than fault slip alone. Comparison with field surveys indicates that the maximum velocity was on the order of 1 m/s in each case. The 2026 Kumamoto earthquake, however, exhibited a more complex velocity history than the other two events. The three earthquakes involved different mechanisms of surface fault displacement, suggesting that these differences may also influence the complexity of the displacement time histories.
\end{abstract}

\section{Introduction}

Co-seismic surface fault displacement provides one of the few direct observational constraints on fault motion during earthquakes. In most cases, such displacement is observed only after fault slip and rupture have ceased, as a static displacement discontinuity; nevertheless, these observations play an important role in constraining both the direction and magnitude of fault slip. When strong-motion seismometers are installed within several hundred meters of a fault, co-seismic displacements in the immediate vicinity of the fault can be obtained \citep[e.g.][]{Fukuyama07, Delouis24}. However, even at distances of only several hundred meters from the fault, the temporal resolution of the fault-slip history that can be reconstructed from seismograms deteriorates \citep{Fukuyama07}, and complementary measurements have been desired for more detailed analyses.

In recent years, several cases have been reported in which closed-circuit television (CCTV) cameras captured the moment of fault motion. During the 2025 Mandalay earthquake ($M_\mathrm{w}$7.7 according to the U.S. Geological Survey [USGS]; see Data and Resources) in Myanmar, video footage recorded the emergence of a strike-slip surface rupture during strong ground motion, followed by approximately 2 s of slip and a final displacement of 2 m. This footage constituted the first video to capture in detail the actual motion of a surface fault and has subsequently been analyzed in numerous studies \citep{Gao25,Hirano25,Kearse25,Latour25}. These studies have extracted a wide range of information, including not only the slip-time history but also the frictional constitutive law on the fault plane \citep{Latour25}, the rupture-propagation velocity at the recording site \citep{Hirano25}, and consistency with theoretically predicted curved fault motion \citep{Kearse25}. Thus, video footage that directly captures fault slip has the potential to provide information unavailable from seismometers, and analyses of the Mandalay earthquake footage established an important precedent.

The Kumamoto earthquake ($M_\mathrm{w}$6.8 according to the USGS; see Data and Resources, or $M_\text{JMA}$7.1\citep{HERP26}) of July 2026 became the second case in which clear surface fault displacement was captured on video. Ten years earlier, during the April 2016 Kumamoto earthquake sequence, the Hinagu and Futagawa faults ruptured successively. In the 2026 event, rupture occurred along the southern Hinagu Fault in a segment that had not ruptured in 2016\citep{HERP26}. The regional topography, with the Yatsushiro Plain on the northwestern side of the fault and mountainous terrain on the southeastern side, suggests a structural setting favorable to uplift of the southeastern side. Indeed, co-seismic slip in the 2026 earthquake comprised both normal-fault and right-lateral strike-slip components. CCTV footage looking eastward from a convenience store shows cracks forming in a rice paddy during strong ground motion, accompanied by uplift of the ground to the east and both upward and right-lateral motion of distant houses. \citet{Toda26} precisely measured the surface fault displacement at the video site. Photograph No. 17 on page 6 of their report corresponds to this location. Its caption, written in Japanese, reads: ``Earthquake fault crossing a rice paddy in Miyaji-machi, Yatsushiro City. Right-lateral displacement of 105 cm and vertical displacement of 90 cm, with the eastern side uplifted. Photograph by Toda, July 31.'' Following a very preliminary report by \citet{Ando26}, more detailed analysis of this footage is expected to provide important constraints on fault motion, as in the case of the 2025 Mandalay earthquake. For that purpose, comparison among the different cases within a unified methodology is required.

In this study, we analyze CCTV footage that captured the emergence of the surface rupture during the 2026 Kumamoto earthquake and reconstruct its temporal evolution. To this end, we tracked the motion of points on the opposite side of the fault after eliminating the effects of strong ground motion. For comparison, we reanalyze the 2025 Mandalay earthquake footage using the same procedure. We also analyze video footage from the February 2018 Hualien earthquake ($M_\mathrm{w}$6.4 according to the USGS; see Data and Resources). Although this footage does not directly record fault slip itself, CCTV cameras captured the site directly above the fault as well as the interior and exterior of a building, allowing the relative displacement between the site and an adjacent building to be investigated. Through these three case studies, we quantify the duration of surface displacement and the complexity of its temporal evolution.

\section{Data and Methods}

On August 14, 2026, a 14-second video was published showing the formation of the surface rupture during strong ground motion caused by the 2026 Kumamoto earthquake (see Data and Resources for the URL).
At 7 seconds into the video, a surface rupture with oblique displacement appeared in the rice paddy, after which the ground and houses on the far side moved toward the upper right of the frame.
We analyzed the video to reconstruct the fault-slip history as follows.

\begin{figure}
\centering
\includegraphics[width=86mm]{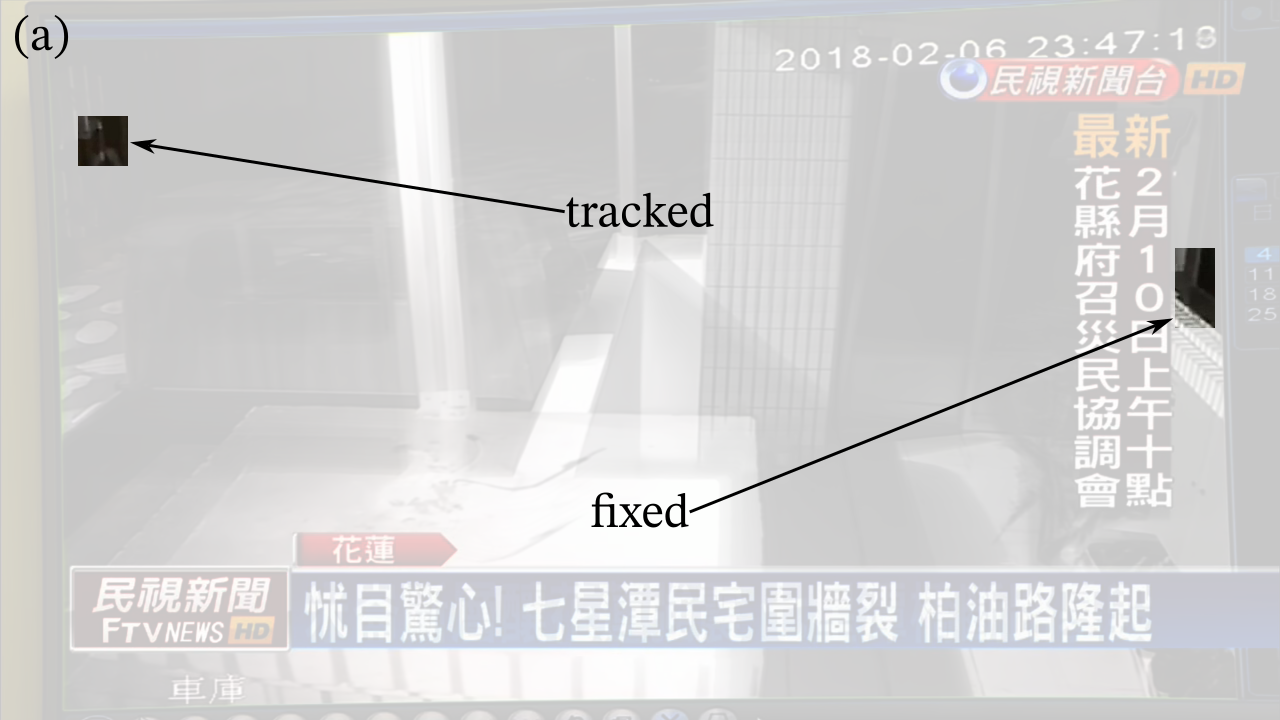}
\includegraphics[width=86mm]{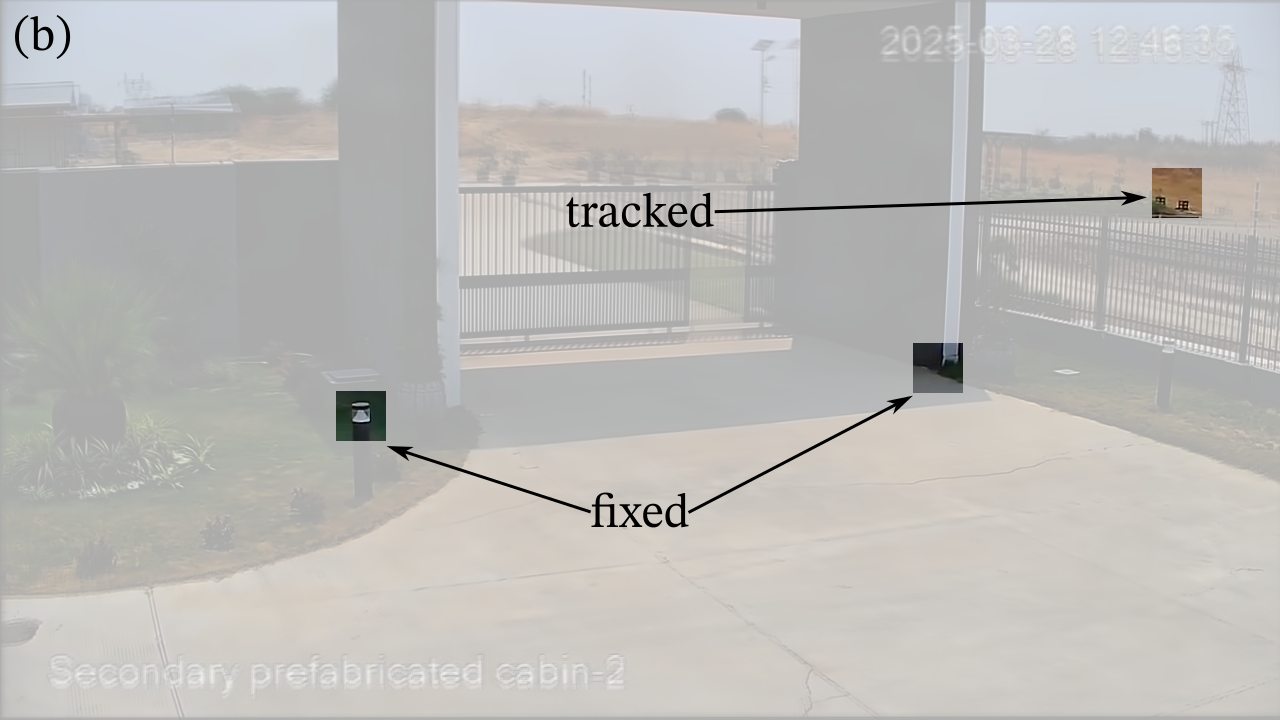}
\includegraphics[width=86mm]{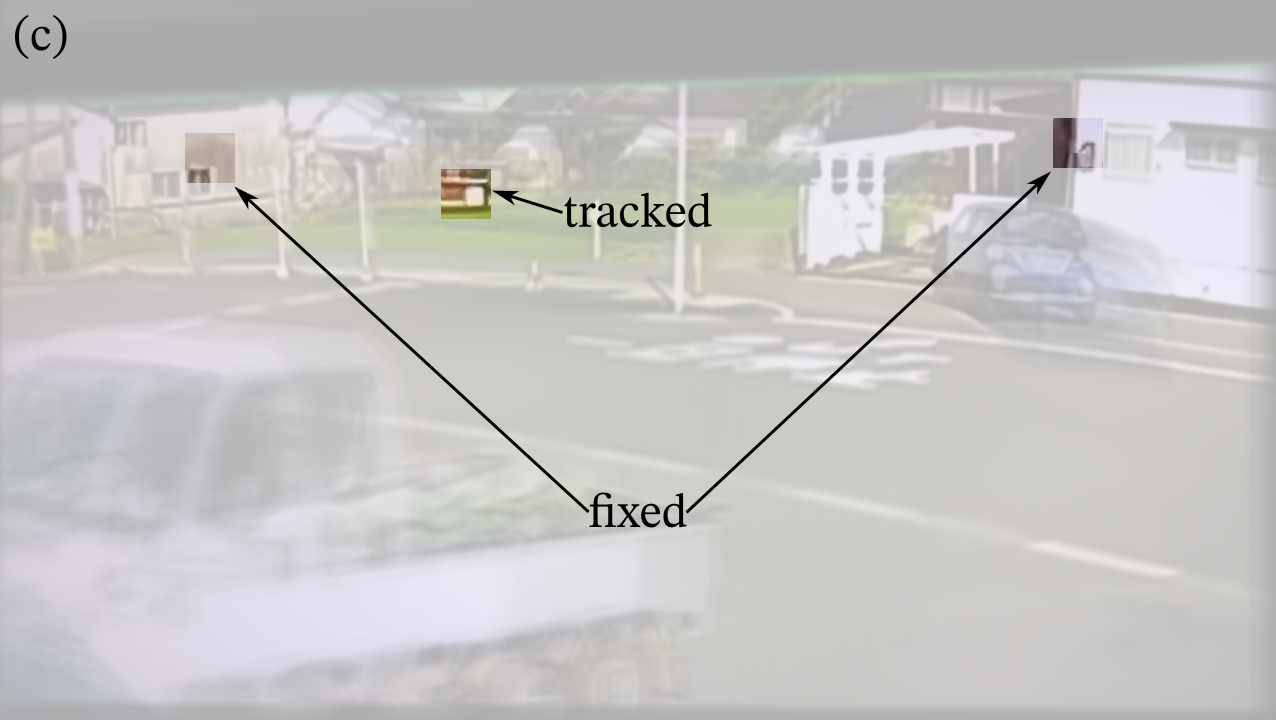}
\caption{\captionwindows}
\label{fig:windows}
\end{figure}

At first inspection, the footage exhibits substantially stronger camera shaking than the Mandalay earthquake footage. Moreover, the publicly available video is not the original recording but a handheld secondary recording of a playback screen, and therefore contains additional camera motion absent from the original footage. To suppress these effects, we extracted two 50$\times$50 pixel windows located on the near side of the fault and identified, in each frame, the coordinates maximizing the correlation with each reference window. All frames were then translated, scaled, and rotated such that these two locations remained fixed (Fig.\ref{fig:windows}). This procedure, based on \citet{Gao25}, reduces the effects of both motion of the original CCTV camera and handheld motion introduced during the secondary recording.

We subsequently extracted a 50$\times$50 pixel window containing part of a structure visible on the far side of the fault and determined, with subpixel precision, the coordinates maximizing the correlation with this window in every frame. These coordinates can be regarded as representing the apparent relative displacement of the ground on opposite sides of the fault plane. Some frames exhibit motion blur caused by strong ground motion. Although precise localization is difficult in such frames, the corresponding correlation coefficients tend to decrease (gray points in the tracking results). The influence of these frames can therefore be reduced in the analysis by assigning lower weights to observations with low correlation coefficients.

\begin{figure}
\centering
\includegraphics[width=86mm]{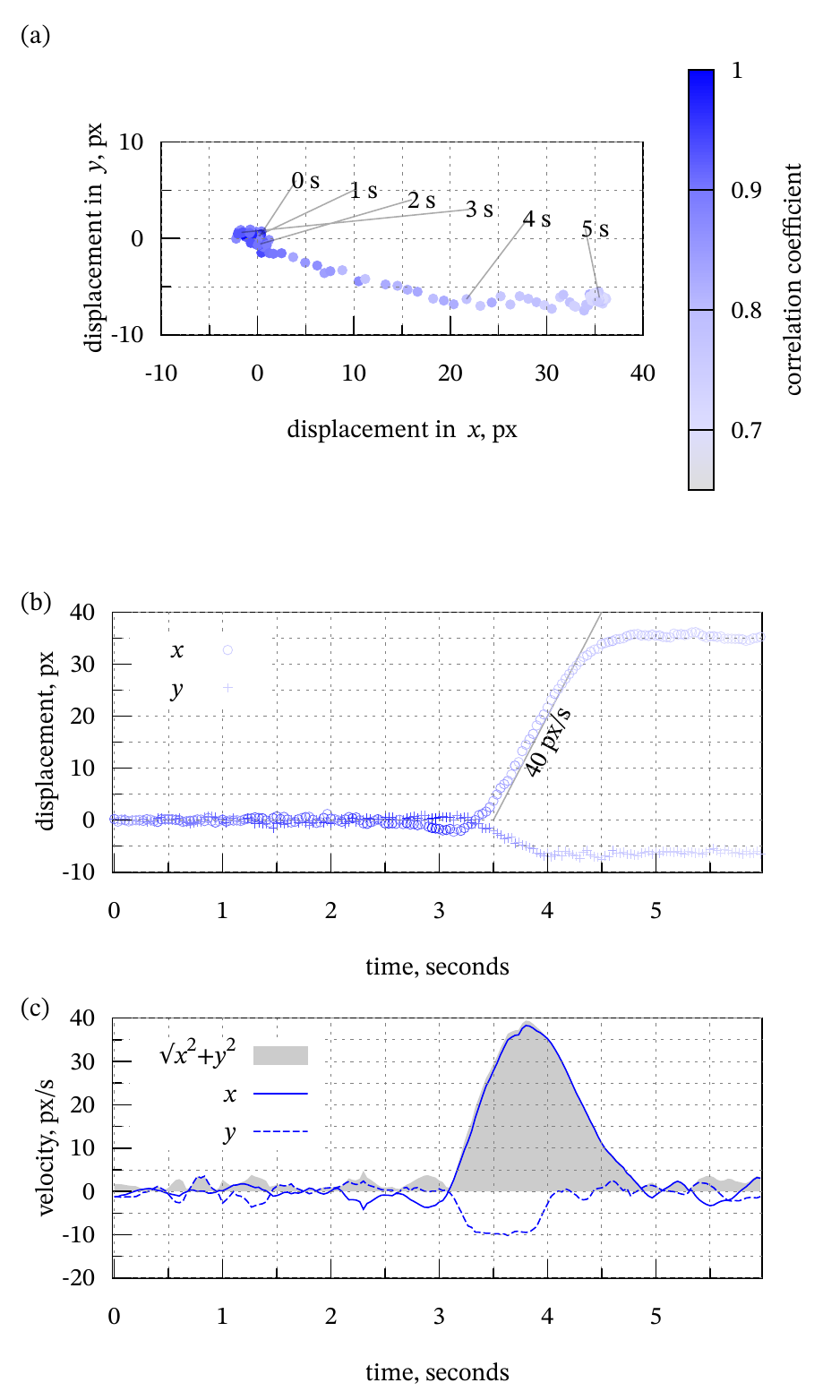}
\caption{\captionMandalay}
\label{fig:Mandalay}
\end{figure}

Once positions have been extracted from each frame, velocity can in principle be obtained by differentiation. Because displacement is initially measured in pixels, the corresponding velocity is represented in pixels per second and is subsequently converted to m/s using field-survey measurements. In practice, simple frame-to-frame differencing of the positions is highly unstable. We therefore applied a Savitzky--Golay-type filter, i.e., weighted local fitting of a second-order polynomial, using the fourth power of the correlation coefficients described above as the weights and using the $n$ frames before and after each time step. Velocity was then obtained by differentiating the fitted polynomial. The Kumamoto and Mandalay videos were recorded at 30 fps, for which we used $n=8$, whereas the Hualien footage had an effective frame rate of 11 fps, for which we used $n=5$. These $n$ values were selected empirically to balance the stability and fidelity of the estimated velocity histories. Specifically, $n$ was kept small enough for the estimated peak velocity to reach the steepest sustained slope evident in each displacement-time plot (straight lines in Figs.~\ref{fig:Hualien}b--\ref{fig:Kumamoto}b); larger values produced excessive smoothing and peak velocities substantially lower than those slopes.

\section{Results}

We first present the results for the Hualien and Mandalay earthquakes as comparative cases, followed by those for the 2026 Kumamoto earthquake.
The values of tracked displacement in pixels and estimated velocity in pixels per second are summarized in Table \ref{tab:measurement}.

\subsection{2018 Hualien Earthquake ($M_\mathrm{w}$6.4)}

Although the Hualien earthquake footage is itself a secondary recording of a playback screen, visual inspection indicates little handheld shaking, rotation, or panning motion throughout the video. We therefore used only one location on the right side of the image as a fixed reference window and tracked a region on the left side (Fig.\ref{fig:windows}a). Additional footage recorded immediately before and after the analyzed sequence from a different viewing angle indicates that the tracked motion corresponds to a building across the road moving closer to the camera-side structure.

\begin{figure}
\centering
\includegraphics[width=86mm]{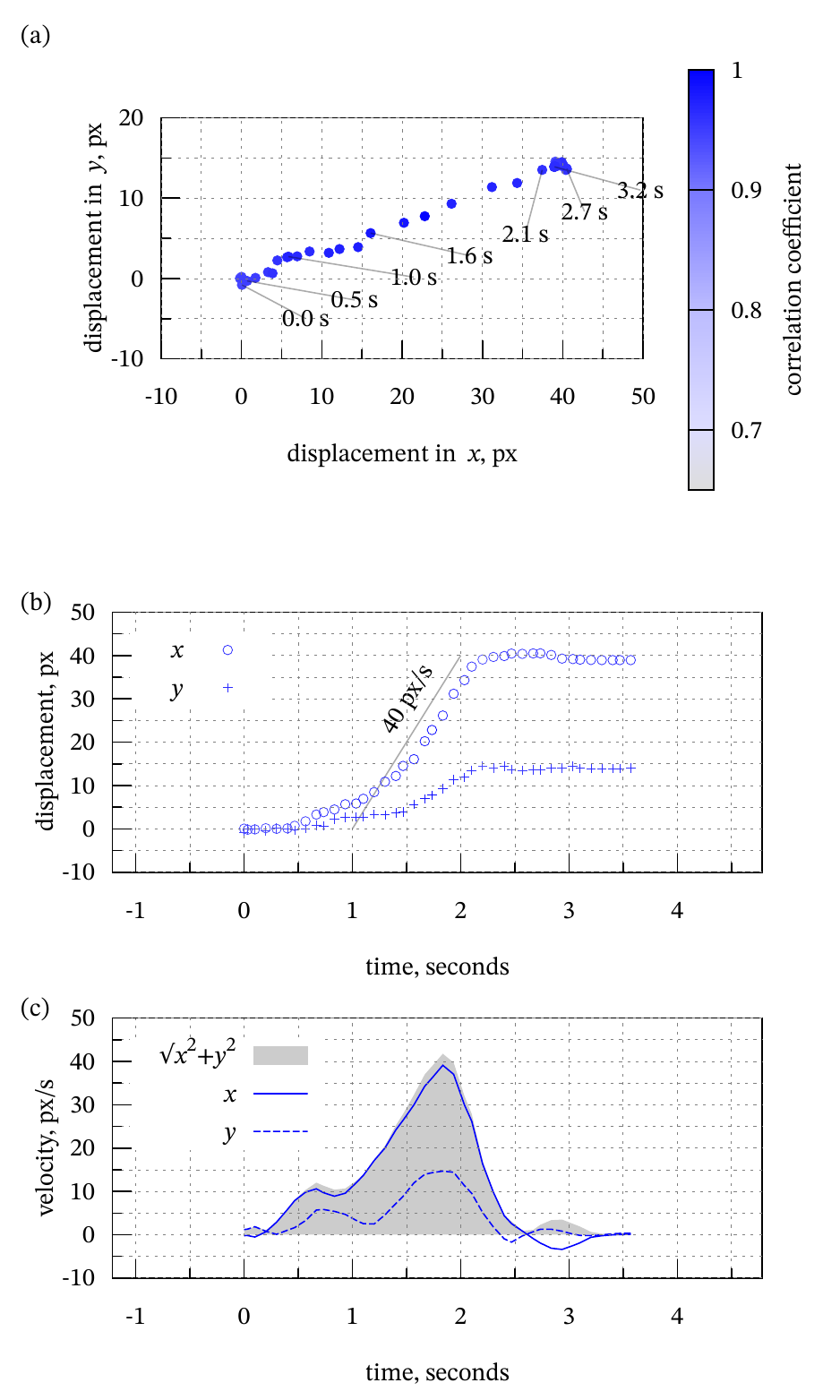}
\caption{\captionHualien}
\label{fig:Hualien}
\end{figure}

The video is short and has a low ($\sim 11$ fps) effective frame rate, but it nevertheless reveals an approximately linear trajectory (Fig.\ref{fig:Hualien}). The tracked motion is completed within approximately 2 s and is broadly unimodal.
Using $n=5$, the velocity estimate in Fig.\ref{fig:Hualien}c reproduces the steepest sustained slope apparent in Fig.\ref{fig:Hualien}b ($\sim 40$ px/s).
A slight velocity pause appears shortly after motion begins, but this feature is comparable to the fluctuation expected from the low effective frame rate and the frame-to-frame scatter of the tracking, and we do not interpret it as a resolved feature of the surface displacement history. As discussed below, this motion likely records not only fault slip but also compression of the ground. Nevertheless, it provides sufficient information for an order-of-magnitude estimate of the duration of surface displacement.

\subsection{2025 Mandalay Earthquake ($M_\mathrm{w}$7.7)}

The Mandalay earthquake footage has been analyzed in several previous studies, all of which obtained mutually consistent results; our analysis yields similarly consistent behavior (Fig.\ref{fig:Mandalay}). Because we apply the same method as for the Hualien and Kumamoto footage, however, the results can be compared directly among the three cases. As noted by \citet{Kearse25}, fault slip during this earthquake did not follow a consistently linear trajectory throughout the slip episode. Near the end of the motion, the velocity in the $y$ direction decreased, producing a slight curvature in the trajectory.

Although this earthquake was substantially larger than the Hualien and Kumamoto events, the duration of slip at the recording site was nearly identical, approximately 2 s. As pointed out by \citet{Hirano25}, the rupture arriving at the recording site underwent transient subshear propagation, and the associated strong ground motion was prominent immediately before fault slip. After suppressing this effect using the present method, we obtain a remarkably smooth displacement-time history.

\subsection{2026 Kumamoto Earthquake ($M_\mathrm{w}$6.8)}

\begin{figure}
\centering
\includegraphics[width=86mm]{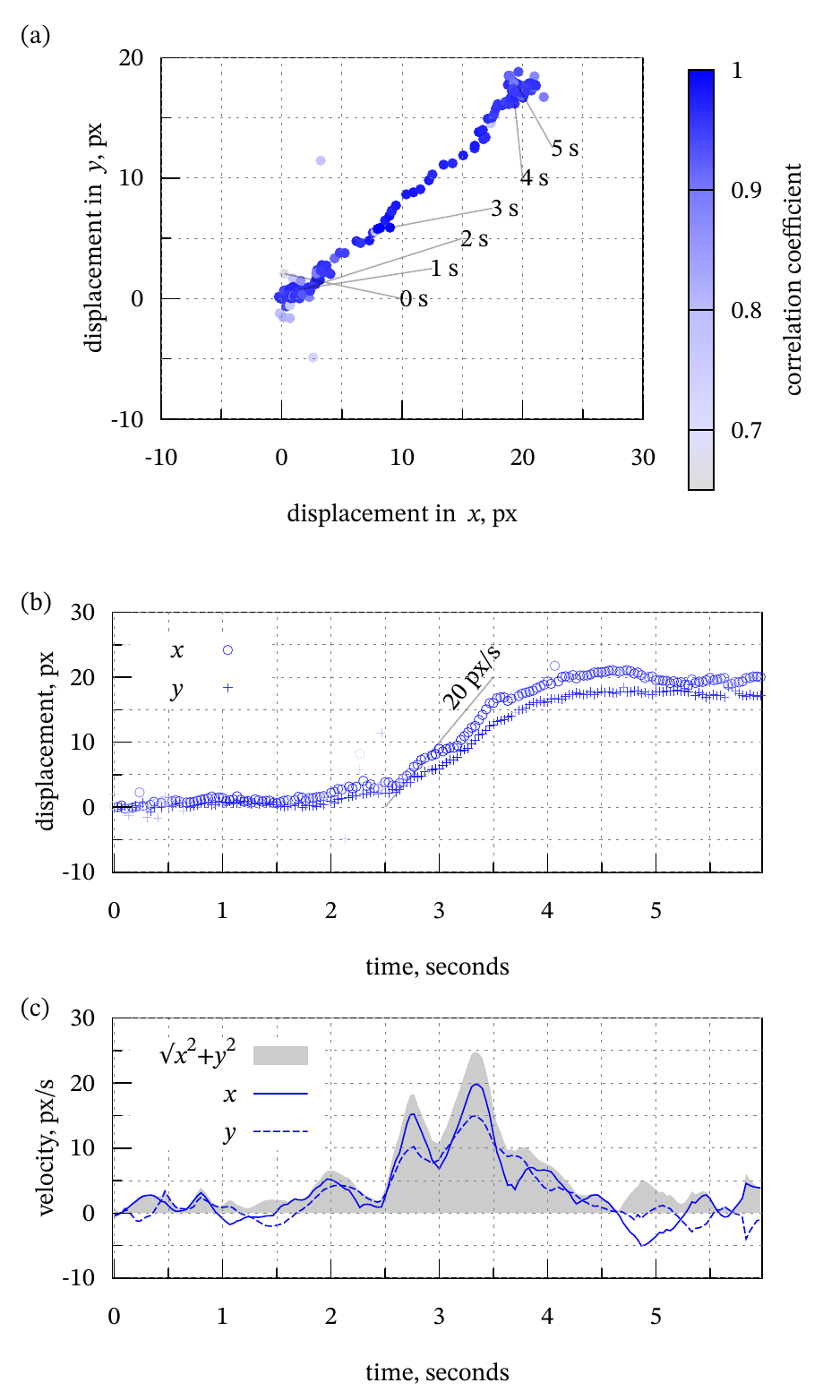}
\caption{\captionKumamoto}
\label{fig:Kumamoto}
\end{figure}

Our results are consistent with those of \citet{Ando26} but resolve the surface fault motion at a temporal resolution approximately an order of magnitude higher (eight measurements over 2.2 s in \citealp{Ando26}). The motion followed an approximately linear trajectory (Fig.\ref{fig:Kumamoto}a), in contrast to the Mandalay case. A slight overshoot is nevertheless observed near the end of the $x$ component in the image (Fig.\ref{fig:Kumamoto}b), which is thought to correspond approximately to horizontal motion. The $x$ and $y$ components in the video exhibit similar displacement amplitudes, consistent with post-earthquake field measurements at the same site of 1.05 m horizontal displacement and 0.90 m vertical displacement \citep{Toda26}.

The slip-velocity history indicates that the motion was largely completed within approximately 2 s of its onset (Fig.\ref{fig:Kumamoto}c). Compared with the Hualien and Mandalay cases, however, the history is highly complicated even aside from the overshoot. Even after the effects of strong ground motion are suppressed as much as possible, the velocity history contains at least three local peaks, at approximately 2.0, 2.7, and 3.3 s. The displacement history includes occasional points that appear to be outliers, but these points have low correlations and therefore exert little influence on the velocity estimate. In fact, the complexity of the velocity history occurs independently of the intervals of low correlation, supporting the interpretation that it reflects complex surface fault motion rather than solely tracking errors associated with motion blur.

\section{Discussion and Conclusions}

A common result among all three cases is that surface motion lasted approximately 2 s. According to USGS source-process analyses (see Data and Resources), the Mandalay earthquake had a rupture duration of 85 s, whereas the Hualien earthquake had a duration of 8 s. Thus, although the USGS whole-fault rupture durations and the local durations measured here are not directly equivalent quantities, the videos show comparable local motion durations for events whose whole-fault durations differ by more than an order of magnitude. According to the direct measurement, the ground produced 0.27 m uplift and 0.80 m compression \citep{Huang19} during the Hualien earthquake; these values reflect not fault slip alone but quite complicated surface deformation, because the area was on the horsetail splay near the northern end of the Milun fault. Therefore, the approach of the adjacent building visible in the video may represent the compression of the ground surface rather than fault slip. Nevertheless, the duration might be a proxy for that of fault slip. For the Kumamoto earthquake, the USGS source-process duration is approximately 16 s (see Data and Resources), again substantially shorter than that of the Mandalay earthquake. Taken together, these three cases show no evident direct scaling between local surface-motion duration and earthquake magnitude.

Where surface displacement is independently known from field surveys, the displacement and velocity obtained in pixels and pixels per second can be converted to meters and meters per second, respectively, by matching the final value of the derived displacement time series to the surveyed displacement. However, the present analysis measures only the $x$- and $y$-direction displacements in the video image, and these components cannot be mapped uniquely onto fault-parallel and vertical components. Here, therefore, we estimate the maximum surface-motion or fault-slip velocity by relating the two-dimensional resultant displacement in the video to the field-survey measurements.
Table \ref{tab:measurement} summarizes the field-measured surface displacements and the maximum velocities inferred from the videos when those measurements are treated as the final displacement.

For the 2018 Hualien earthquake, 0.27 m of vertical displacement and 0.80 m of horizontal compression were identified within the recording site \citep{Huang19}, whereas the strike-slip component is unknown. The video displacement also clearly contains shortening in the fault-normal direction, which is interpreted as the compression. The final displacement visible in the video is therefore likely to correspond to $\sqrt{0.27^2+0.80^2} \sim 0.84$ m. Nevertheless, if this motion does not directly represent the amount of fault slip, its velocity may provide an order-of-magnitude estimate of the fault-slip velocity. On this basis, taking the resultant of the measured uplift and shortening as the final displacement, the peak velocity is estimated to be at least 0.84 m/s (Table \ref{tab:measurement}). This value is the smallest of the three cases discussed below, although the difference may reflect the limitation that the Hualien footage records complicated surface displacement rather than fault slip directly.

For the 2025 Mandalay earthquake, we obtain the largest velocity among the three cases, at least 2.1 m/s. The field-survey report \citep{Gao25} does not provide a vertical-displacement measurement, but the photographs indicate that the vertical component is clearly small relative to the horizontal displacement; the inferred velocity is therefore considered robust. The persistence of a relatively high slip velocity over a duration comparable to those of the other cases accounts for the large surface displacement. Satellite-image analysis, however, indicates a broader displacement discontinuity of approximately 4 m \citep[e.g.,][]{Hirano25}. One possible explanation for this discrepancy is that displacement resolvable in satellite imagery occurs not only on the principal slip plane but also across a finite-width deformation zone surrounding the fault. If so, the fault slip visible in the video does not account for all of the shallow deformation. Under the simple interpretation that approximately 2 m of the total 4 m displacement is accommodated on the principal slip plane, the difference would constrain how shallow deformation is partitioned between the principal slip plane and the surrounding deformation zone.

For the 2026 Kumamoto earthquake, field surveys reported both horizontal and vertical displacement discontinuities\citep{Toda26}. If these are assumed to correspond to the $x$ and $y$ displacements in the video, respectively, the resultant velocity is approximately 1.3 m/s, of the same order as that of the Mandalay earthquake. Notably, laboratory rock-friction experiments and observations of many natural earthquakes suggest that peak slip velocities inferred from seismic or in-situ elastic waves are in the range of 1--8 m/s \citep{McGarr03}. The results for both the Mandalay and Kumamoto earthquakes fall within this range. The estimate of \citet{McGarr03} represents the maximum slip velocity on the fault plane weighted by slip and does not account for surface displacement; it should instead be regarded as a slip velocity associated with a seismologically resolvable centroid. It is therefore nontrivial that the surface values obtained here are of the same order, and the mechanism responsible for this similarity warrants further investigation.

A particularly notable feature of the Kumamoto earthquake is the complexity of its slip history. The strongly fluctuating velocity function obtained even after minimizing the effects of strong ground motion and motion blur in the video suggests that relatively compliant near-surface materials do not necessarily undergo purely dissipative, stable sliding. Unlike the Mandalay case, the Hinagu Fault dips at approximately 75${}^\circ$ and accommodates both strike-slip and normal-fault components. In addition, rupture was not a simple, horizontally propagating pulse but also propagated in the depth direction. These factors may have contributed to the observed complexity.

In this study, we analyzed three videos that captured co-seismic surface displacement. Despite differences in earthquake magnitude, the inferred displacement durations are of the same order, whereas the complexity of the displacement histories differs substantially among the cases. As the number of CCTV cameras in operation continues to increase, similar recordings are likely to become more common. Video analysis can therefore provide observations complementary to those obtained from seismometers and field surveys.

\section*{Data and Resources}
    The URLs of the videos are as follows:
    \begin{itemize}
        \item Hualien: \url{https://www.youtube.com/watch?v=XH_KBw_D6jo}
        \item Mandalay: \url{https://www.youtube.com/watch?v=77ubC4bcgRM}
        \item Kumamoto: \url{https://www.threads.com/@m.m.s.k.a.ao_matsumura/post/Db-dELDE6mc}
    \end{itemize}
    The USGS event pages used for earthquake magnitudes and source-process information are
    \url{https://earthquake.usgs.gov/earthquakes/eventpage/us1000chhc/executive} for the 2018 Hualien earthquake,
    \url{https://earthquake.usgs.gov/earthquakes/eventpage/us7000pn9s/executive} for the 2025 Mandalay earthquake, and
    \url{https://earthquake.usgs.gov/earthquakes/eventpage/us6000tgb9/executive} for the 2026 Kumamoto earthquake.
    The author believes that using these videos in this manuscript is consistent with the principles of fair use under U.S. copyright law for the purposes of scholarship and research (\url{https://www.copyright.gov/fair-use/}).
    A preliminary version of the results included in this manuscript (Figs.\ref{fig:windows} and \ref{fig:Kumamoto}) was already published by the author on social media: \url{https://twitter.com/Bimaterial/status/2088183164279878003}.
    All links in this paragraph were last accessed on 2026-08-16.

\section*{Declaration of Competing Interests}
    The author declares no competing interests.

\section*{Acknowledgments}
    The author is sincerely grateful to the uploaders of the three videos.

\bibliographystyle{apalike-ejor}
\bibliography{mybibfile}

\clearpage
\begin{table}
\begin{center}
\caption{Estimated peak velocities calibrated using reference displacements measured at the three video sites. The field measurements listed under ``Displacement (measurement)'' are from \citet{Huang19} for the 2018 Hualien earthquake, \citet{Gao25} for the 2025 Mandalay earthquake, and \citet{Toda26} for the 2026 Kumamoto earthquake. N/A denotes not available. The motion in the 2018 Hualien case does not represent pure fault slip; see the text for details. The $x$ and $y$ values of the video-tracking displacement are defined as the difference between the average of the tracked coordinates over the first five frames and that over the last five frames of the analyzed sequence.}
\begin{tabular}{rc|rrr}
                &             & 2018 Hualien    & 2025 Mandalay & 2026 Kumamoto   \\ \hline
Displacement    & Vertical    & $0.27$ m        & N/A           & $0.90$ m        \\
(measurement)   & Strike-slip & N/A             & $\ge 1.93$ m  & $1.05$ m        \\
                & Compression & $0.80$ m        & N/A           & N/A             \\
                & Total       & $\ge 0.84$ m    & $\ge 1.93$ m  & $\ge 1.38$ m    \\ \hline
Displacement    & $x$         & $39.6$ px       & $34.9$ px     & $20.0$ px       \\
(video tracking)& $y$         & $-14.1$ px      &  $6.3$ px     & $-17.0$ px      \\
                & Total       & $42.1$ px       & $35.5$ px     & $26.2$  px      \\ \hline
Peak velocity   & Total       & $41.8$ px/s     & $39.4$ px/s   & $\ge 24.8$ px/s \\
(estimation)    & Total       & $\ge 0.84$ m/s  & $\ge 2.1$ m/s & $\ge 1.3$ m/s
\end{tabular}
\label{tab:measurement}
\end{center}
\end{table}

\clearpage

\end{document}